# The Cloud Adoption Toolkit: Addressing the Challenges of Cloud Adoption in the Enterprise

Ali Khajeh-Hosseini, David Greenwood, James W. Smith, Ian Sommerville

Cloud Computing Co-laboratory, School of Computer Science, University of St Andrews, UK
{akh, dsg22, jws7, ifs}@cs.st-andrews.ac.uk

**Abstract.** Cloud computing promises a radical shift in the provisioning of computing resource within the enterprise. This paper: i) describes the challenges that decision makers face when attempting to determine the feasibility of the adoption of cloud computing in their organisations; ii) illustrates a lack of existing work to address the feasibility challenges of cloud adoption in the enterprise; iii) introduces the Cloud Adoption Toolkit that provides a framework to support decision makers in identifying their concerns, and matching these concerns to appropriate tools/techniques that can be used to address them. The paper adopts a position paper methodology such that case study evidence is provided, where available, to support claims. We conclude that the Cloud Adoption Toolkit, whilst still under development, shows signs that it is a useful tool for decision makers as it helps address the feasibility challenges of cloud adoption in the enterprise.

## 1 Introduction

Cloud computing is the latest effort in delivering computing resources as a service. It represents a shift away from computing as a product that is owned, to computing as a service that is delivered to consumers over the internet from large-scale data centres – or 'clouds'. Cloud computing is currently being exploited by technology start-ups due to its marketed properties of scalability, reliability and cost-effectiveness. Enterprises are also beginning to show an interest in cloud computing due to these promised benefits, however at present much ambiguity and uncertainty exists regarding the actual realisability of these promised benefits in the enterprise. Whilst there is much hype surrounding cloud computing, particularly around its cost savings which are based on simplistic assumptions, we believe the technology is still likely to have a profound effect on the ways software will be procured, developed and deployed, similar to the effect of moving from mainframes to PCs.

This paper's original contribution is to propose the Cloud Adoption Toolkit, which provides a collection of tools that can be used to support decision making during the adoption of cloud computing in the enterprise. Our toolkit is based on a framework to organise thinking about decision makers' concerns and match these to tools that address these concerns, where each tool enables decision makers to focus on and model different attributes of their organisations or IT systems. These models can then be used to reason about and investigate cloud adoption decisions. For example, by modelling a system's hardware infrastructure and applications, it becomes possible to estimate the costs of running that system in a cloud, and hence decide whether deploying that system in the cloud would be cost effective. Furthermore, by identifying the impacts of a proposed system to people's work activities, the proposed system's practical and socio-political feasibility can be determined. For example, a system may be cost effective yet socio-politically infeasible if it potentially decreases job satisfaction and undermines existing power bases or organisational values [1].

This paper is structured as follows: Section 2 introduces cloud computing and discusses the importance of modelling in systems engineering; Section 3 describes the challenges of cloud adoption in the enterprise and illustrates the challenges by looking at a case study; Section 4 describes the conceptual framework behind the Cloud Adoption Toolkit and provides details of its individual tools; and Section 5 concludes that the toolkit is promising and presents our future work.

## 2      Background

### 2.1    Cloud Computing

There are many definitions of cloud computing [e.g. 2, 3, 4]. The US National Institute of Standards and Technology (NIST) has published a working definition that has captured the commonly agreed aspects of cloud computing; it defines cloud computing as "a model for enabling convenient, on-demand network access to a shared pool of configurable computing resources (e.g., networks, servers, storage, applications, and services) that can be rapidly provisioned and released with minimal management effort or service provider interaction" [5]. The NIST definition describes cloud computing as being composed of:
- Five characteristics: on-demand self-service, broad network access, resource pooling, rapid elasticity, and measured service.
- Four deployment models: private clouds, community clouds, public clouds, and hybrid clouds.
- Three service models: Software as a Service (SaaS), Platform as a Service (PaaS), and Infrastructure as a Service (IaaS).

Figure 1 provides an overview of the common deployment and service models in cloud computing, where the three service models could be deployed on top of any of the four deployment models.

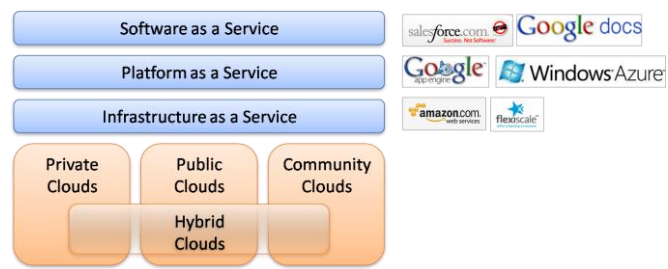

Fig. 1. Cloud computing deployment and service models

## 2.2     Modelling

Making decisions regarding implementations of enterprise IT systems is complicated due to the intricate interrelationships between people and technology that decision makers must contend with. Successful systems are a consequence of a stabilised mesh of stakeholder interests, values, know-how and technological characteristics [6, 7]. Modelling has an important role to play in the engineering of successful enterprise systems because it enables decision makers to make sense of and represent these intricate interrelationships [8, 9].

Successful models address two challenges; the representational challenge and the informational challenge. The representation challenge is concerned with how to best represent the information included in the model to make it easy to understand. The informational challenge comprises what to leave-in and what to leave-out of a model. The informational challenge can be extremely challenging in the field of systems engineering as some systems such as social systems are hard to define and delimit. In order to address the informational challenge of complex systems, frameworks or paradigms can be employed to reduce complexity by introducing simplifying assumptions.

Modelling IT infrastructure is particularly challenging as enterprise IT environments have many variables and these variables have multiple interactions such that knowledge of their causal intricacies is especially important for selecting appropriate modelling primitives. One important way of managing this modelling challenge is by using frameworks, based upon rigorous studies of the dynamics of the domain and experiential knowledge where rigorous studies are unavailable, to select appropriate primitives [6, 7].

## 3     Challenges of Cloud Adoption

Cloud computing is not simply about a technological improvement of data centres but a fundamental change in how IT is provisioned and used [11]. Enterprises need to consider the benefits, risks and the effects of cloud computing on their organizations and usage-practices in order to make decisions about its adoption and use [12]. In the enterprise, the "adoption of cloud computing is as much dependent on the maturity of organisational and cultural (including legislative) processes as the technology, per se" [13]. The adoption of cloud computing is not going to happen overnight – some predict that it could take between 10 to 15 years before the typical enterprise makes this shift [14]. Therefore we are currently at the start of a transition period during which many decisions need to be made with respect to cloud adoption in the enterprise.

Cloud adoption decisions are challenging because of a range of practical and socio-political reasons. It is unlikely that all organisations will completely outsource their back-end computing requirements to a cloud service provider. Rather, they will establish heterogeneous computing environments based on dedicated servers, organisational clouds and possibly more than one public cloud provider. How their application portfolio is distributed across this environment depends not just on technical issues but also on socio-technical factors (e.g., concerns about costs, confidentiality, and control), the impact on work practices and constraints derived from existing business models. Therefore, the challenges that a cloud adoption toolkit must address are: i) to provide accurate information on costs of cloud adoption; ii) to support risk management; and iii) to ensure that decision makers can make informed trade-offs between the benefits and risks.

It is important that decision makers take into account these considerations because the use of cloud computing has significant implications for an enterprise as a whole [1]. For example, we recently performed a feasibility analysis of a proposed cloud-based IT system at an SME in the oil and gas industry [1]. We found that despite the promised financial benefits, opportunities to remove tedious work from IT staff and the potential to enter new marketplaces, almost all of the stakeholder groups were neutral or reluctant to support a move to the cloud due to concerns regarding its impact on their work, increased risk of dependence upon third parties and its implications for customer service and support.

A recent review of the academic research done in cloud computing revealed that there are currently no mature techniques or toolkits available to support decision making during the adoption of cloud computing in the enterprise [12, 15]. In industry, [16] and [17] provide examples of typical offerings from IT consultancies that attempt to fill this gap. Such approaches have two problems: they are based on closed proprietary tools that are not widely available; and they are often accompanied by expensive consultancy periods. In contrast, we argue that given the Cloud Adoption Toolkit, enterprises can assess the feasibility of using cloud computing in their organisations themselves. However, the toolkit can also be used by decision makers to verify the claims made by IT consultancies and cloud service providers.

## 4      The Cloud Adoption Toolkit

The Cloud Adoption Toolkit comprises a conceptual framework for organizing decision makers' concerns and matching these to tools that address them. Decision makers can use any tools/techniques that they wish to, however, we provide five tools/techniques that we believe to be extremely useful: Technology Suitability Analysis; Cost Modelling; Energy Consumption Analysis; Stakeholder Impact Analysis; and Responsibility Modelling.

### 4.1      Conceptual Framework

The purpose of the conceptual framework is to organise decision makers' thinking about the concerns that they and other stakeholders have, and the tools that can be used to explore these concerns. It is important that decision makers view the proposed cloud adoption project from multiple stakeholders' perspectives in order to learn from a diverse range of stakeholder concerns and receive a broad range of feedback from the organisational environment. Figure 2 provides an overview of the Cloud Adoption

Toolkit and how it can be used. Decision makers would start with a Technology Suitability Analysis, and if the cloud is found to be suitable for their system, they would proceed by investigating either the costs of running the system on public clouds, or the energy consumption (and hence energy costs) of running the system on private clouds. At the same time, Stakeholder Impact Analysis can be performed to assess the impacts of using cloud computing on the work of stakeholders in the enterprise. If these analyses show that running the system on the cloud is a viable option, then Responsibility Modelling can be performed to identify and analyse the risks associated with the operation of the system on the cloud, where different cloud providers could be responsible for different aspects of the system.

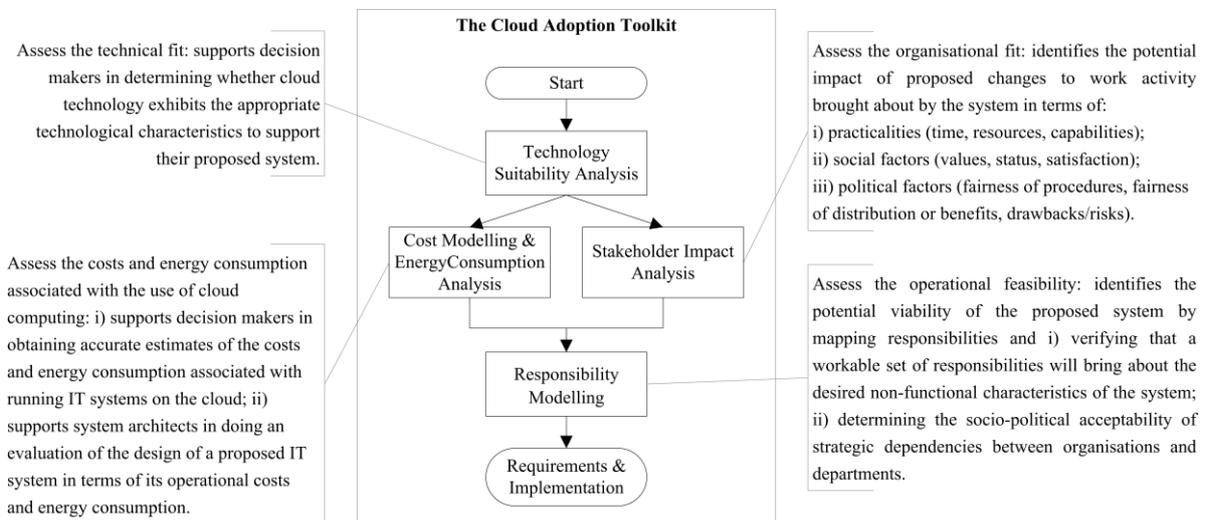

**Fig. 1.** Cloud adoption conceptual framework

### 4.2    Technology Suitability Analysis

The purpose of Technology Suitability Analysis is to support decision makers in determining whether cloud computing exhibits the appropriate technological characteristics to support their proposed system. Understanding the characteristics of cloud computing is extremely important as it has the potential to exhibit radically different properties to those of traditional enterprise data centres. This is mainly due to the cloud's highly scalable nature, physical resource sharing between virtual machines, potential issues to do with communication over the internet and insufficient guarantees regarding the up-time and reliability of processing and data storage services. For example, typical IaaS offerings make no reassuring guarantees about server uptime or network performance which has important implications for the viability of certain classes of software architectures and business critical systems.

The Technology Suitability Analysis comprises a simple checklist of questions to provide a rapid assessment of the potent suitability of a particular cloud service for a specific enterprise IT system. The checklist is still under development but the current version, shown in Table 1, analyses eight characteristics and quickly provides an indication of the cloud's suitability for a proposed IT system.

**Table 1.** Technological Suitability Analysis

| Desired Technology Characteristic | Questions |
|---|---|
| 1. Elasticity | - Does your software architecture support scaling out?<br>- If not, will scaling up to a bigger server suffice? |
| 2. Communications | - Is the bandwidth within the cloud and between the cloud and other systems sufficient for your application?<br>- Is latency of data transfer to the cloud acceptable? |
| 3. Processing | - Is the CPU power of instances appropriate for your application at the expected operating load?<br>- Do server instances have enough memory for your application? |
| 4. Access to hardware / bespoke hardware | - Does your cloud provider provide the required access to hardware components or bespoke hardware? |
| 5. Availability / dependability | - Does your cloud provider provide an appropriate SLA?<br>- Are you able to create the appropriate availability by mixing geographical locations or service providers? |
| 6. Security requirements | - Does your cloud service provider meet your security requirements? (e.g. do they support multi-factor authentication or encrypted data transfer) |
| 7. Data confidentiality and privacy | - Does your cloud provider provide sufficient data confidentiality and privacy guarantees? |
| 8. Regulatory requirements | - Does your cloud provider comply with the required regulatory requirements of your organisation? |

### 4.3   Cost Modelling

The purpose of the Cost Modelling tool is to support cloud adoption decisions in two different ways:

1. To support decision makers in obtaining accurate cost estimates of running IT systems on the cloud. The tool helps decision makers investigate the costs of migrating an existing IT system or deploying a new IT system on the cloud, the costs of migrating an IT system from one cloud to another, or even future costs based upon predictions of future workload.
2. To support system architects in evaluating the design of a proposed IT system with respect to its operational costs, with the aim of minimizing the costs.

Simplistically, IT system procurement is based on obtaining estimates for a proposed system, then getting those estimates signed-off by management or the client to allow the procurement to proceed. Capital and operational budgets are often kept separate in many organizations, and procurement costs have to be known in advance before approval can be gained. Currently, the estimation process is based on predicting the maximum quantity of resources (i.e. processing power, memory, storage etc.) that a system might need and provisioning at that level. However, most of the acquired resources remain unused during normal operation as the estimates were based on peak load. In fact recent figures show server utilization in traditional data centres ranging from 5% to 20% [2].

The utility billing model of cloud computing has a certain degree of uncertainty that goes against current procurement policies. The uncertainty relates to: i) the actual resources consumed by a system, which are determined by its load; ii) the deployment option used by a system, which can affect its costs as resources like bandwidth are more expensive between clouds compared to bandwidth within clouds; iii) the cloud service provider's pricing model, which can change at any time. The utility billing model is also a shift away from capital to operational budgeting, and many enterprises are less savvy about operational budgeting for IT than they are for capital budgeting.

Cost Modelling extends the capabilities of UML deployment diagrams [8], which enable a system's deployment to be modelled. In its essence, a UML deployment diagram enables users to model the deployment of software artefacts onto hardware nodes. The Cost Modelling tool enables users to model a system's software applications and how they could be deployed on cloud, traditional or hybrid infrastructures. The model is then processed to give users an accurate estimate of the operational costs of their system. The models can take into account future resource demands therefore enabling for situations where traditional infrastructure may not initially be cost-effective, yet will become cost effective with future workload increases.

The concepts behind the Cost Modelling tool were used in a recent case study that compared the infrastructure costs of deploying a system on Amazon's Elastic Compute Cloud (EC2) versus a traditional data centre. The tool showed that the system infrastructure in the case study would have cost around 37% less over 5 years on EC2 compared with a traditional data centre [1].

### 4.4    Energy Consumption Analysis

The purpose of Energy Consumption Analysis is to support decision making regarding the optimum energy usage of IT resources when a system is deployed on a private cloud. This tool will help to inform decision making and aid the design of cloud-based architectures by enabling the assessment of potential trade-offs between energy efficiency and performance in order to build systems to particular requirements. This tool is currently under development and investigations into this area are ongoing. However, the idea of reducing a system's energy consumption is promising as a traditional blade server consumes 50% of its peak energy when it is switched on and doing only 10% of its capable work. This disproportionate amount of consumption is caused by the server's baseline consumption [18]. If a server is not being fully utilized then the baseline consumption will dominate the overall energy consumption of a system. Cloud computing provides an alternative to this situation as Infrastructure as a Service providers such as Amazon deploy multiple virtual servers on the same physical server. For example, on the StACC Private Cloud (www.cs.st-andrews.ac.uk/stacc) we use a mapping of 8 virtual servers to one physical machine.

Virtualisation techniques enable systems to achieve far higher levels of utilisation, and therefore can save energy by reducing the number of physical servers that need to be operating.

### 4.5    Stakeholder Impact Analysis

The purpose of Stakeholder Impact Analysis is to support decision makers in determining the socio-political viability, or benefits and risks, of a proposed IT system. This is important as cloud adoption projects are not merely technological upgrades but involve the reconfiguration of working practices and technologies to take full advantage of the benefits offered by the technology [11]. The socio-political benefits and risks associated with a proposed IT system are determined by identifying the impact of changes to stakeholders' work activities in terms of their practicalities (time, resources, and capabilities), social factors (interests, values, status, and satisfaction) and political factors (their perception of the fairness of decision making procedures and the distribution of benefits, drawbacks and risks). This information enables decision makers to make a judgement about the risk that specific stakeholders will hold unsupportive attitudes towards the proposed system and therefore indicates the overall socio-political feasibility of the system.

Stakeholder Impact Analysis has been successfully used to support decision making regarding the feasibility of migrating a client-server application (a data acquisition and quality monitoring system) to Amazon EC2 [1]. The analysis revealed that the proposed cloud migration would have many implications for the organization including non-technical areas such as the finance and marketing departments. Overall a positive net benefit was perceived from the perspectives of the business development functions of the enterprise and the more junior levels of the IT support functions. A zero net benefit was perceived by the project management and support management functions of the enterprise and a negative net benefit was perceived by the technical manager and the support engineer functions of the enterprise.

The analysis identified numerous potential benefits and risks associated with the migration. Most notably, opportunities for improved cash flow management, opportunities to offer new products/services, and removal of tedious work were identified as benefits. In contrast the following notable risks were also elicited: the deterioration of customer care and service quality; increased dependency on external $3^{rd}$ parties; and departmental down-sizing. Overall the decision makers decided that for this particular situation the benefits/opportunities did not outweigh the drawbacks/risks and therefore decided a migration was not viable in the short term.

### 4.6    Responsibility Modelling

The purpose of responsibility modelling is to support decision makers in determining the operational viability of a proposed IT system. Responsibility modelling also helps decision makers in identifying and analysing risks associated with the operation of complex IT systems [19]. Responsibility modelling is particularly important for systems deployed on the cloud as the responsibilities for constructing, operating, maintaining, and managing the system can be divided across multiple organisations, departments and cloud service providers, and therefore identifying and managing the risks associated with the discharge of responsibilities is important to the operational viability of the IT system.

The viability of the system is determined by: i) identifying the set of responsibilities that must be discharged for the system to operate according to a set of non-functional requirements; ii) who is responsible for what; iii) whether the configuration of responsibilities is likely to meet non-functional requirements of the system; and iv) determining the practical, social and political viability of the discharge of responsibilities so that the system exhibits appropriate non-functional characteristics e.g. up-time, responsiveness, resilience, maintainability and recoverability.

## 5    Conclusion

Cloud computing is currently being exploited by technology start-ups due to its marketed properties of scalability, reliability and cost effectiveness. Enterprises are also beginning to show an interest in cloud computing due to these promised benefits, however at present much ambiguity and uncertainty exists regarding the actual realisablility of these promised benefits in the enterprise.

This paper demonstrated that current feasibility approaches fall short in terms of enabling enterprise decision makers to determine the viability of using cloud computing, and that our Cloud Adoption Toolkit offers a promising starting point. Our conclusions are limited due to position paper methodology that this paper uses and its reliance upon limited case study data.

The Cloud Adoption Toolkit is designed such that it can incorporate new tools to account for emerging factors with respect to cloud computing, and in the future we aim to accommodate other industrial concerns such as how cloud computing can reduce an organisation's carbon emissions. We are currently in the process of planning a number of industrial case studies to evaluate and further develop the Cloud Adoption Toolkit.

**Acknowledgements**

We thank the Scottish Informatics and Computer Science Alliance (SICSA) and the EPSRC for funding the authors. We also thank our colleagues at the UK's Large-Scale Complex IT Systems Initiative (www.lscits.org) for their comments.